\documentclass[reprint,
 amsmath,amssymb,
 aps,
]{revtex4-2}

\usepackage{graphicx}
\usepackage{dcolumn}
\usepackage{bm}
\usepackage{xcolor}

\begin{document}


\title{The space-time Talbot effect}

\author{Layton A. Hall$^{1}$}
\author{Murat Yessenov$^{1}$}
\author{Sergey A. Ponomarenko$^{2,3}$}
\author{Ayman F. Abouraddy$^{1}$}
\affiliation{$^{1}$CREOL, The College of Optics \& Photonics, University of Central~Florida, Orlando, FL 32816, USA}
\affiliation{$^{2}$Department of Electrical and Computer Engineering, Dalhousie University, Halifax, Nova Scotia B3J 2X4, Canada}%
\affiliation{$^{3}$Department of Physics and Atmospheric Science, Dalhousie University, Halifax, Nova Scotia B3H 4R2, Canada}%

\begin{abstract}
The Talbot effect, epitomized by periodic revivals of a freely evolving periodic field structure, has been observed with waves of diverse physical nature in space and separately in time, whereby diffraction underlies the former and dispersion the latter. To date, a combined spatio-temporal Talbot effect has not been realized in any wave field because diffraction and dispersion are independent physical phenomena, typically unfolding at incommensurable length scales. Here we report the observation of an optical ‘space-time’ Talbot effect, whereby a spatio-temporal optical lattice structure undergoes periodic revivals after suffering the impact of both diffraction and dispersion. The discovered space-time revivals are governed by a single self-imaging length scale, which encompasses both spatial and temporal degrees of freedom. Key to this effect is the identification of a unique pulsed optical field structure, which we refer to as a V-wave, that is endowed with intrinsically equal diffraction and dispersion lengths in free space, thereby enabling self-imaging to proceed in lockstep in space and time.
\end{abstract}

\maketitle

Since Talbot reported in 1836 his observation of axial self-imaging in an optical field endowed with a periodic transverse spatial profile \cite{Talbot36PM}, the eponymous effect has remained a source of fascination \cite{Rayleigh81PM,Montgomery67JOSA,Berry96JMO}. Such a periodic field first undergoes diffractive spreading in space such that the initial periodic structure is progressively lost, only for the original periodic profile to be revived at planes separated by the Talbot distance $z_{\mathrm{T},x}$. In optics, the Talbot effect has found myriad applications \cite{Wen13AOP}, ranging from structured illumination in fluorescence microscopy \cite{Chowdhury18arxiv} to prime-number decomposition \cite{Pelka18OE} and phase-locking of laser arrays \cite{Tradonsky17AO}. An analogous \textit{temporal} Talbot effect exists where dispersive spreading in time replaces diffractive spreading in space. In this scenario, pulses in a periodic train first spread and overlap temporally, followed by subsequent revivals of the original structure in the time domain at planes separated by the temporal Talbot distance $z_{\mathrm{T},t}$. The temporal Talbot effect has been observed thus far in single-mode optical fibers \cite{Jannson81JOSA,Andrekson93OL}, where it has been employed in removing distortion \cite{Azana99AO}, filtering noise \cite{Fernandez04JOSAB}, and pulse-rate multiplication \cite{Arahira98JLT,Shake98EL,Atkins03IEEE}.

To date, the spatial and temporal Talbot effects have been studied as separate physical phenomena. Can Talbot self-imaging be observed simultaneously in space \textit{and} time? in other words, starting with a field having a spatio-temporal periodic lattice structure, can spontaneous revivals occur after the lattice disintegrates under the joint impact of diffraction in space and dispersion in time? It is easy to recognize that combined spatio-temporal self-imaging is in fact possible if independent revivals are arranged to coincide $z_{\mathrm{T},x}=z_{\mathrm{T},t}$; which requires that the dispersion length \textit{precisely} match the diffraction length (or an integer multiple thereof). Indeed, even slightly mismatched diffraction and dispersion lengths can obscure such coinciding self-imaging revivals. However, this requirement is in conflict with the fundamental fact that diffraction and dispersion are physically independent phenomena that are governed by altogether different parameters of the field, and thus usually operate on disparate length scales. Consequently, the axial self-imaging lengths for the spatial and temporal Talbot effects are usually incommensurate (for example, $z_{\mathrm{T},t}\!\gg\!z_{\mathrm{T},x}$ for optical waves). This creates challenging technical hurdles, and a combined spatio-temporal Talbot effect has \textit{not} been realized thus far.

Here we propose and demonstrate an optical field configuration that enables the observation of a `space-time' (ST) Talbot effect; that is, Talbot self-imaging unfolding jointly in space and time at a \textit{single} axial length scale underlying both diffraction and dispersion. Rather than simply combining decoupled spatial and temporal Talbot effects at coinciding planes, we observe a new Talbot effect in which the field evolution in space and time are correlated by virtue of the field structure itself. The key to realizing this ST Talbot effect is the identification of a novel pulsed optical beam structure -- dubbed a `V-wave' -- whose spatial and temporal degrees of freedom are inextricably linked. V-waves are endowed with a unique spectral structure whereby each spatial frequency is precisely associated with a single prescribed temporal frequency (or wavelength). The spatial and temporal frequencies are selected to be linearly proportional to each other, leading to a V-shaped spatio-temporal spectrum that occupies a reduced-dimensionality sub-space with respect to that of conventional optical wave packets whose spectra are separable with respect to the spatial and temporal degrees of freedom \cite{SalehBook07}. This field structure endows the axial wave number with a `Janus' form (Janus is the two-faced Roman god): when expanded in terms of the spatial frequencies, the axial wave number describes diffractive spreading in space; but when expanded in terms of the temporal frequencies, it describes dispersive spreading in time. Therefore, the spectral structure of a V-wave results in group velocity dispersion (GVD) in free space in concurrence with diffraction. Crucially, the length scales associated with diffraction and dispersion of a V-wave are guaranteed to be intrinsically matched.

By discretizing the spatio-temporal spectrum of a V-wave, the optical field forms a periodically checkered spatio-temporal lattice. We observe rich axial dynamics enfolding in space and time as this lattice propagates freely. The lattice first degenerates into a disordered pattern after experiencing diffractive and dispersive spreading, prior to the emergence of subsequent periodic revivals. Because diffraction and dispersion of the underlying V-wave unfold in lockstep, a \textit{single} self-imaging length scale, the ST Talbot length $z_{\mathrm{ST}}$, dictates the planes at which revivals take place jointly in space and time. Remarkably, there is no need for adjusting the decoupled spatial and temporal degrees of freedom. Indeed, the field invariably maintains equal diffraction and dispersion lengths even when spatial or temporal parameters are varied. The observation of the ST Talbot effect thus becomes viable in free space over experimentally convenient length scales \textit{without} recourse to a highly dispersive medium.

Our experiments unveil unique features jointly in space-time that are associated with the purely spatial and temporal Talbot effects. For example, we record the displacement of the spatio-temporal lattice by half a period along both space and time at mid-Talbot planes, and we record a spatio-temporal lattice-period halving (rate-doubling) at quarter-Talbot planes. Finally, the tight association between the spatial and temporal spectral degrees of freedom leads to a new manifestation of the `veiled' Talbot effect \cite{Yessenov20arXiv}: despite the dramatic evolution of the spatio-temporal field structure with propagation, these changes are nevertheless altogether masked in the time-averaged intensity where \textit{no} axial dynamics are displayed. The intensity in space features instead a diffraction-free transverse periodic structure, having a period half that of the underlying spatio-temporal lattice structure, whose axial dynamics are hidden from view. Our results may potentially impact high-speed optical-beam steering \cite{Shaltout19Science}, three-dimensional nonlinear lithography \cite{Maruo08LPR,Kumi10LC}, and clock synchronization \cite{Bhaduri20NP}, all of which can benefit from the controllable synthesis of spatio-temporal optical lattices.

\section{Conventional Talbot effects in space and time}

To set the stage for elucidating the concept of the ST Talbot effect, we first briefly describe the purely spatial and temporal optical Talbot effects, in addition to assessing the prospect of observing them simultaneously (see Supplementary for details).

\subsection{The spatial Talbot effect}

\begin{figure}[t!]
\centering 
\includegraphics[width=7.6cm]{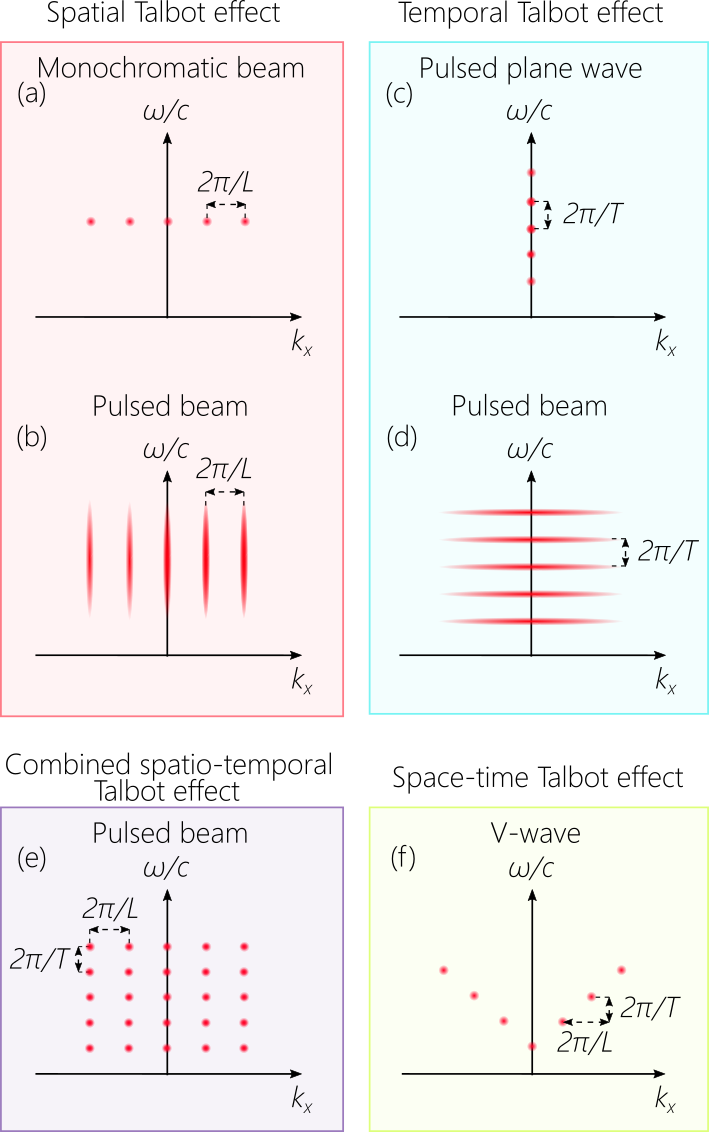} 
\caption{\textbf{Spectral representation of Talbot effects in space, time, and space-time.} (a) In the purely \textit{spatial} Talbot effect, the spatial spectrum of a monochromatic beam represented in $(k_{x},\tfrac{\omega}{c})$-space is sampled periodically along $k_{x}$ at multiples of $\tfrac{2\pi}{L}$. (b) The spatial Talbot effect when implemented with a pulsed beam. (c) In the purely \textit{temporal} Talbot effect, the spectrum of a plane-wave pulse is sampled periodically along $\omega$ at multiples of $\tfrac{2\pi}{T}$. (d) The temporal Talbot effect when implemented with a pulsed beam. (e) Combined -- but independent -- spatial and temporal Talbot effects, in which the spatial and temporal spectra are sampled periodically at multiples of $\tfrac{2\pi}{L}$ and $\tfrac{2\pi}{T}$, respectively. The spatio-temporal spectra in (a-e) are all separable with respect to $k_{x}$ and $\omega$. (f) The ST Talbot effect that makes use of a pulsed beam having a V-shaped spatio-temporal spectrum, $\omega=\omega_{\mathrm{o}}+\alpha c|k_{x}|$. The spatial spectrum is sampled at multiples of $\tfrac{2\pi}{L}$, which corresponds to sampling the temporal spectrum at multiples of  $\tfrac{2\pi}{T}$, where $L=\alpha cT$.}
\label{fig:classification}
\end{figure}

The field of a monochromatic beam at a temporal frequency $\omega_{\mathrm{o}}$ can be written as $E(x,z;t)\!=\!e^{i(k_{\mathrm{o}}z-\omega_{\mathrm{o}}t)}\psi(x,z)$, where $k_{\mathrm{o}}\!=\!\tfrac{\omega_{\mathrm{o}}}{c}$ is the associated wave number, $c$ is the speed of light in vacuum, and the spatial envelope $\psi(x,z)$ in the paraxial regime is expressed as the angular spectrum:
\begin{equation}\label{Eq:Spatial}
\psi(x,z)=\int\!dk_{x}\,\widetilde{\psi}_{x}(k_{x})\,\exp{\{ik_{x}x\}}\,\exp{\{-i\tfrac{k_{x}^{2}}{2k_{\mathrm{o}}}z\}};
\end{equation}
where the spatial spectrum $\widetilde{\psi}_{x}(k_{x})$ is the Fourier transform of $\psi(x,0)$, we assume, for simplicity, that the field is uniform along the transverse coordinate $y$, and we retain the transverse and axial coordinates $x$ and $z$ with wave-vector components $k_{x}$ and $k_{z}$, respectively (we refer to $k_{x}$ as the spatial frequency). If the initial transverse spatial profile $\psi(x,0)$ is periodic with period $L$, the spatial spectrum is discretized at $k_{x}\!=\!m\tfrac{2\pi}{L}$, where $m$ is an integer index, as illustrated in Fig.~\ref{fig:classification}(a). Upon discretization of the spatial spectrum, the phase term $\exp{\{-i\tfrac{k_{x}^{2}}{2k_{\mathrm{o}}}z\}}$ responsible for diffractive spreading takes the form $\exp{\{-i2\pi m^{2}z/z_{\mathrm{T},x}\}}$. The field first diffracts before the initial profile is revived $\psi(x,\ell z_{\mathrm{T},x})\!=\!\psi(x,0)$ at axial planes separated by the spatial Talbot distance $z_{\mathrm{T},x}\!=\!\tfrac{2L^{2}}{\lambda_{\mathrm{o}}}$ ($\ell$ is an integer and $k_{\mathrm{o}}=\tfrac{2\pi}{\lambda_{\mathrm{o}}}$) \cite{Wen13AOP}. This Talbot effect can still be observed with pulsed or broadband fields whose spectrum is separable with respect to $k_{x}$ and $\omega$ [Fig.~\ref{fig:classification}(b)], when space-time coupling is minimal $\Delta\omega\!\ll\!\omega_{\mathrm{o}}$, or after removing chromatic aberrations \cite{Packros84OC,Lancis95JMO,Guerineau00OC}. 

\subsection{The Temporal Talbot effect}

When a plane-wave optical pulse $E(x,z;t)\!=\!e^{i(nk_{\mathrm{o}}z-\omega_{\mathrm{o}}t)}\psi(z,t)$ propagates in a dispersive medium, the temporal envelope $\psi(z,t)$ expressed as an angular spectrum is:
\begin{equation}\label{Eq:Temporal}
\psi(z,t)=\int\!d\Omega\,\widetilde{\psi}_{t}(\Omega)\exp{\{-i\Omega(t-\tfrac{z}{\widetilde{v}})\}}\exp{\{i\tfrac{1}{2}k_{2}\Omega^{2}z\}};
\end{equation}
where the temporal spectrum $\widetilde{\psi}_{t}(\Omega)$ is the Fourier transform of $\psi(0,t)$, $n$ is the refractive index at the central frequency $\omega_{\mathrm{o}}$, $\Omega\!=\!\omega-\omega_{\mathrm{o}}$ is the temporal frequency with respect to $\omega_{\mathrm{o}}$, $\widetilde{v}$ is the group velocity $\tfrac{1}{\widetilde{v}}=\tfrac{dk_{z}}{d\Omega}\big|_{\Omega=0}$, and $k_{2}=\tfrac{d^{2}k_{z}}{d\Omega^{2}}\big|_{\Omega=0}$ is the GVD parameter \cite{SalehBook07}. Because dispersive spreading in time is the analog of diffractive spreading in space \cite{Kolner94IEEEJQE,vanHowe06JLT}, a temporal Talbot effect occurs when the field profile at $z\!=\!0$ is periodic in time with period $T$. The temporal spectrum of this pulse train is discretized at $\omega\!=\!m\tfrac{2\pi}{T}$ [Fig.~\ref{fig:classification}(c)], and the phase term $\exp{\{i\tfrac{1}{2}k_{2}\Omega^{2}z\}}$ responsible for dispersive spreading takes the form $\exp{\{i2\pi m^{2}z/z_{\mathrm{T},t}\}}$. The pulse train first disperses and the pulses overlap temporally before it reconstitutes itself axially at planes separated by the temporal Talbot distance $z_{\mathrm{T},t}\!=\!\tfrac{T^{2}}{\pi|k_{2}|}$; $\psi(\ell z_{\mathrm{T},t},t)=\psi(0,t-\ell\tfrac{z_{\mathrm{T},t}}{c})$. Because plane-wave pulses do not experience GVD in free space, this effect requires a dispersive medium for its observation. 

To date, the temporal Talbot effect has \textit{not} been realized in a freely propagating optical field, and has been observed only in single-mode fibers. In such a fiber ($k_{2}\approx-25$~ps$^{2}$/km at $\lambda_{\mathrm{o}}\sim1.5$~$\mu$m), $z_{\mathrm{T},t}>10$~km for a pulse train at a repetition rate of 1~GHz. Reducing this length requires either increasing the repetition rate or increasing the fiber dispersion (e.g., by utilizing a fiber Bragg grating \cite{Azana99AO,longhi00OL}). Whereas $z_{\mathrm{T},x}$ is governed purely by the field structure, $z_{\mathrm{T},t}$ is governed by factors that are extraneous to the optical field: the GVD offered by available optical materials and technological limits on realizable repetition rates of laser pulse trains \cite{Jannson81JOSA,Andrekson93OL,Mitschke98OPN,Fatome04OC}. Note that the spatio-temporal spectrum for a pulsed guided mode is approximately separable with respect to $k_{x}$ and $\omega$, and realizing a guided periodic pulse train entails discretization along only $\omega$ [Fig.~\ref{fig:classification}(d)].

\subsection{The combined spatio-temporal Talbot effect}

After considering the spatial and temporal Talbot effects separately, we inquire whether they can be observed \textit{simultaneously} in an optical field having the form of a spatio-temporal lattice (periodic in time \textit{and} space). If such a field propagates freely in a dispersive medium, can the lattice undergo axial revivals, so that the initially periodic spatio-temporal structure reemerges spontaneously after both diffraction \textit{and} dispersion?

\begin{figure*}[ht!]
\centering 
\includegraphics[width=16cm]{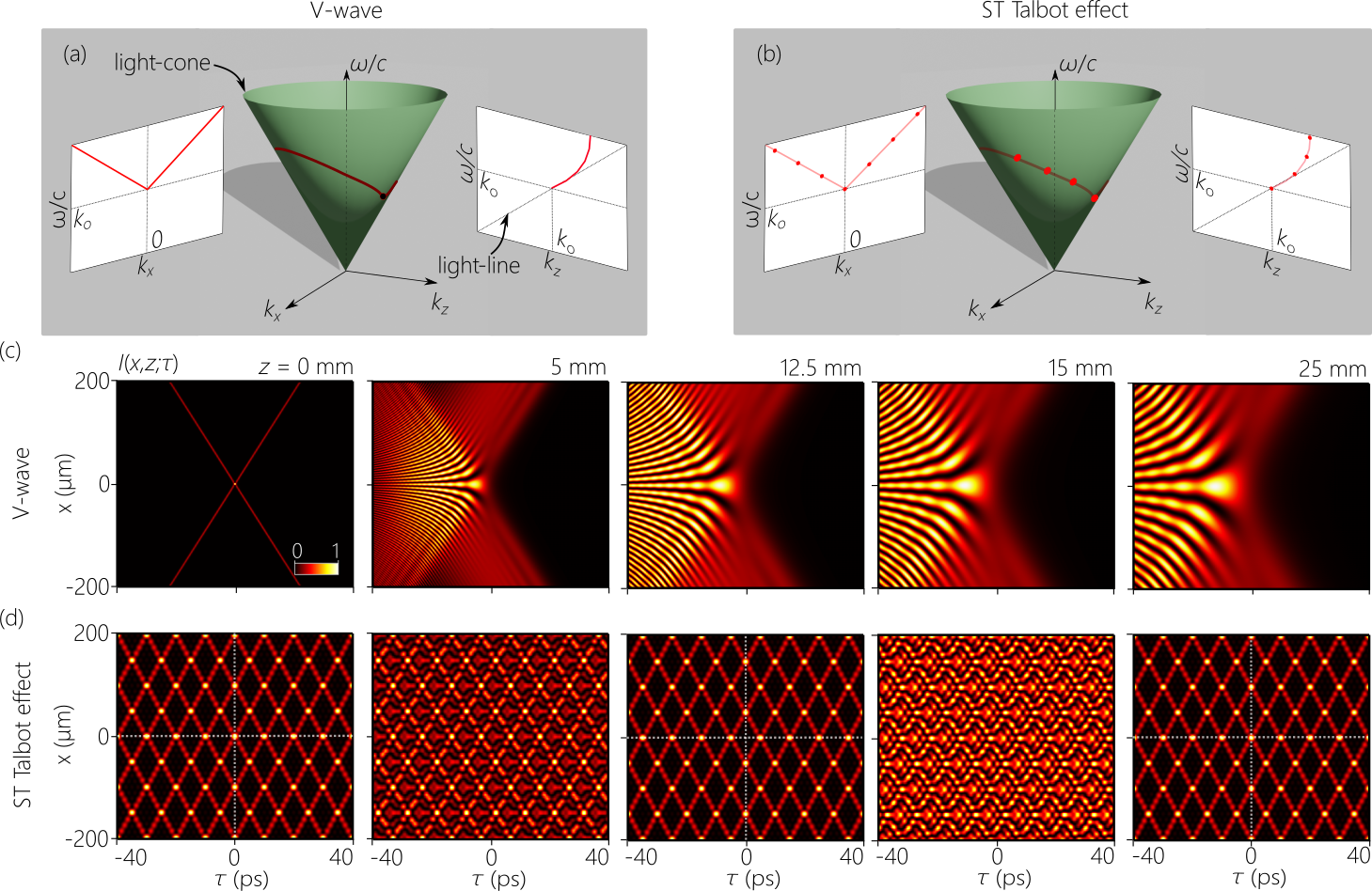} 
\caption{\textbf{Concept of the ST Talbot effect.} (a) Representation of the spectral support domain of a V-wave on the surface of the light-cone and its spectral projections onto the $(k_{x},\tfrac{\omega}{c})$ and $(k_{z},\tfrac{\omega}{c})$ planes. (b) The ST Talbot effect is realized by discretizing the spatio-temporal spectrum of the V-wave in (a). (c) Spatio-temporal profiles $I(x,z;\tau)$ of a V-wave propagating in free space in a reference frame moving at $c$ and experiencing GVD. The profiles are calculated at axial planes $z\!=\!0$, 5, 12.5, 15, 25~mm using the parameters $\alpha\!=\!0.033$, $\lambda_{\mathrm{o}}\!=\!800$~nm, and temporal bandwidth is $\Delta\lambda\approx2$~nm. Experiments confirm these predictions as shown in Supplementary Figure~6. (d) Spatio-temporal intensity profiles $I(x,z;\tau)$ corresponding to the spatio-temporal spectrum shown in (b) and demonstrating the ST Talbot effect. The profiles make use of the same parameters from (c) and are calculated at $z\!=\!0$, 5, 12.5, 15, 25~mm, corresponding to 0,$0.2z_{\mathrm{ST}}$, $0.5z_{\mathrm{ST}}$, $0.6z_{\mathrm{ST}}$ and $z_{\mathrm{ST}}$. The spatial spectrum is sampled along $k_{x}$ at multiples of $\tfrac{2\pi}{L}$ with $L=100$~$\mu$m, resulting in $T=10$~ps and $z_{\mathrm{ST}}=25$~mm. Despite the dramatic differences between the evolution of the field in (c) and (d), their spatio-temporal spectra are identical except for the discretization of the latter. The white dotted lines are guides for the eye and highlight the half-period shift in space and time at the mid-Talbot plane ($z=0.5z_{\mathrm{ST}}$) with respect to the full-Talbot planes ($z=0$ and $z=z_{\mathrm{ST}}$).}
\label{fig:theory}
\end{figure*}

Starting with the field of a paraxial pulsed beam traveling in a dispersive medium $E(x,z;t)=e^{i(nk_{\mathrm{o}}z-\omega_{\mathrm{o}}t)}\psi(x,z;t)$, whose spatial and temporal degrees of freedom are separable, the envelope $\psi(x,z;t)$ can be separated approximately into a product $\psi(x,z;t)\approx\psi_{x}(x,z)\psi_{t}(t,z)$, where $\psi_{x}(x,z)$ and $\psi_{t}(t,z)$ are given by Eq.~\ref{Eq:Spatial} and Eq.~\ref{Eq:Temporal}, respectively -- provided the spectrum is sufficiently narrow to avoid space-time coupling (Supplementary). The spatial envelope $\psi_{x}(x,z)$ undergoes diffractive spatial spreading, while the temporal envelope $\psi_{t}(t,z)$ experiences dispersive temporal spreading. Imposing on the field a periodic transverse spatial profile (period $L$) \textit{and} a periodic temporal profile (period $T$) entails sampling its spatio-temporal spectrum along $k_{x}$ and $\omega$ at multiples of $\tfrac{2\pi}{L}$ and $\tfrac{2\pi}{T}$, respectively [Fig.~\ref{fig:classification}(e)], thereby combining the discretized spatial and temporal spectra from Fig.~\ref{fig:classification}(b) and Fig.~\ref{fig:classification}(d), respectively. If $z_{\mathrm{T},x}\!=\!z_{\mathrm{T},t}$ (or $z_{\mathrm{T},x}\!=\!z_{\mathrm{T},t}/N$, $N$ is an integer), then coinciding spatial and temporal Talbot effects are manifested in the reemergence of the initial spatio-temporal profile at the common Talbot planes.

To date, simultaneous revivals in space and time have \textit{not} been observed using any physical wave. In the case of optical waves, for example, the prospect for realizing such an effect is marred by the fundamental incommensurability of the spatial and temporal Talbot effects ($z_{\mathrm{T},t}\gg z_{\mathrm{T},x}$), which arises from the fundamental mismatch between the diffraction and dispersion length scales. Equalizing $z_{\mathrm{T},x}$ and $z_{\mathrm{T},t}$ for a freely propagating field in a bulk dispersive medium runs afoul of the constraints outlined above. Furthermore, a difference between $z_{\mathrm{T},x}$ and $z_{\mathrm{T},t}$ of even a few percent can obscure the targeted effect (Supplementary Figure~1), which has thus far precluded observing combined spatial and temporal optical Talbot effects.


\section{The space-time Talbot effect}

In light of these challenges, realizing the optical Talbot effect in space \textit{and} time requires an altogether different strategy. We proceed to show that adopting a particular class of optical fields -- for which we coin the name `V-waves' -- whose spatio-temporal spectra are \textit{not} separable with respect to $k_{x}$ and $\omega$, which helps circumvent the fundamental incommensurability of the diffraction and dispersion lengths. By virtue of the angular dispersion introduced into their spectral structure, V-waves have two unique characteristics in free space when contrasted with conventional optical wave packets: first, V-waves can experience extremely large GVD in absence of a dispersive medium; and, second, the diffraction and dispersion lengths are identical. Once the spatio-temporal spectrum of a V-wave is discretized along $k_{x}$ and $\omega$ [Fig.~\ref{fig:classification}(f)], so that the optical field is initially periodic in $x$ and $t$, then these unique characteristics combine to enable the realization of a `space-time' (ST) Talbot effect over small distances in free space. Rather than independent spatial and temporal Talbot lengths, the lattice evolution is characterized by a \textit{single} Talbot length scale in both space and time. 

To bring out the unique structure of V-waves, we start by noting that the angular spectrum of any optical field (Eq.~\ref{Eq:Spatial} and Eq.~\ref{Eq:Temporal}) contains a phase factor $e^{ik_{z}z}$ responsible for the axial dynamics. For a pulsed paraxial beam (Eq.~\ref{Eq:Spatial}) in free space:
\begin{equation}\label{Eq:DiffractionPhase}
k_{z}\,\approx\,k_{\mathrm{o}}+\frac{\Omega}{c}-\frac{k_{x}^{2}}{2k_{\mathrm{o}}},
\end{equation}
where the last term is responsible for diffractive spreading. If instead we consider a plane-wave pulse in a dispersive medium (Eq.~\ref{Eq:Temporal}), then:
\begin{equation}\label{Eq:DispersionPhase}
k_{z}\,\approx\,nk_{\mathrm{o}}+\frac{\Omega}{\widetilde{v}}+\frac{1}{2}k_{2}\Omega^{2},
\end{equation}
where the last term is responsible for dispersive pulse spreading.

Is it possible to sculpt the spatio-temporal spectrum of an optical field in free space so that the expansion of $k_{z}$ takes on \textit{both} forms given in Eq.~\ref{Eq:DiffractionPhase} and Eq.~\ref{Eq:DispersionPhase} simultaneously? In other words, we aim to identify an optical field structure in which $k_{z}$ has a `Janus' form: when expressed in terms of $k_{x}$ it takes on the form in Eq.~\ref{Eq:DiffractionPhase} (diffraction), but when expressed in terms of $\Omega$ it takes on the form in Eq.~\ref{Eq:DispersionPhase} (dispersion)? Such a feature requires that $\Omega$ and $k_{x}$ be proportional to each other:
\begin{equation}\label{Eq:VSpectrum}
\Omega=\Omega(k_{x})=\alpha c|k_{x}|,
\end{equation}
which involves introducing angular dispersion into the field structure (Supplementary). The dimensionless proportionality constant $\alpha$ is:
\begin{equation}
\alpha=\frac{1}{c\sqrt{-k_{2}k_{\mathrm{o}}}},
\end{equation}
which combines the parameters governing diffraction ($\lambda_{\mathrm{o}}$) and dispersion ($k_{2}$), and the GVD here is anomalous ($k_{2}$ is negative-valued) \cite{SalehBook07}. In other words, the spatio-temporal spectrum of the sought-after field must be endowed with a precise structure in which each spatial frequency $\pm k_{x}$ is associated with a single temporal frequency $\omega$ corresponding to the V-shaped spectrum in Eq.~\ref{Eq:VSpectrum}, which we thus call a `V-wave' [Fig.~\ref{fig:theory}(a)]. Because the axial wave number $k_{z}$ of a V-wave can be expanded either in the form in Eq.~\ref{Eq:DiffractionPhase} or that in Eq.~\ref{Eq:DispersionPhase}, the diffraction and dispersion lengths are intrinsically one and the same. A freely propagating V-wave in free space ($n=1$) therefore undergoes diffractive spreading in space \textit{and} dispersive spreading in time, with both phenomena proceeding in lockstep [Fig.~\ref{fig:theory}(c) and Supplementary Movie~1 and Supplementary Movie~2]. Furthermore, combining the constraint in Eq.~\ref{Eq:VSpectrum} with the free-space dispersion relationship $k_{x}^{2}+k_{z}^{2}=(\tfrac{\omega}{c})^{2}$, the group velocity $\widetilde{v}$ of the V-wave is: $\tfrac{1}{\widetilde{v}}=\tfrac{dk_{z}}{d\Omega}\big|_{\Omega=0}=\tfrac{1}{c}$, which guarantees that the second terms in Eqs.~\ref{Eq:DiffractionPhase} and \ref{Eq:DispersionPhase} are also equal. Crucially for our purposes here, the magnitude of the GVD parameter $|k_{2}|=\tfrac{1}{\alpha^{2}c^{2}k_{\mathrm{o}}}$ can be tuned to extremely high values unavailable in known optical materials. For example, a readily realizable value of $\alpha=0.025$ at $\lambda_{\mathrm{o}}\sim1$~$\mu$m produces $|k_{2}|\sim10^{5}\times$ that of silica, thus dramatically reducing the dispersion length even when utilizing narrow spectral linewidths.

Imposing a periodic spatial structure of period $L$ at $z=0$ on a V-wave implies discretization of its spatial spectrum along $k_{x}$ at multiples of $\tfrac{2\pi}{L}$. In conventional pulsed fields with separable spatio-temporal spectra [Fig.~\ref{fig:classification}(a)-(e)], discretization of $k_{x}$ does \textit{not} affect the temporal profile. In contrast, because $\omega$ is linearly related to $k_{x}$ in a V-wave (Eq.~\ref{Eq:VSpectrum}), discretizing along $k_{x}$ results in a concomitant discretization along $\omega$ at multiples of $\tfrac{2\pi}{T}$, $\omega\!=\!\omega_{\mathrm{o}}+\tfrac{2\pi}{T}|m|$ [Fig.~\ref{fig:classification}(f) and Fig.~\ref{fig:theory}(b)], with the spatial period $L$ and temporal period $T$ related through
\begin{equation}\label{Eq:LandT}
L=\alpha cT,
\end{equation}
so that the discretized angular spectrum then takes the form: 
\begin{equation}
\psi(x,z;t)=\sum_{m}\widetilde{\psi}_{m}\,\,e^{i2\pi mx/L}\,\,e^{-i2\pi|m|(t-z/c)/T}\,\,e^{i2\pi m^{2}z/z_{\mathrm{ST}}}.
\end{equation}
That is, periodic modulation of the transverse spatial profile of a V-wave simultaneously imprints a periodic pulse-train structure in the time domain, and $\alpha$ determines the proportionality between the spatial and temporal lattice periods. The last phase term $\exp{\{i2\pi m^{2}z/z_{\mathrm{ST}}\}}$ applies to diffractive spreading in space \textit{and} dispersive spreading in time, \textit{both} of which are now dictated by a single ST Talbot length scale,
\begin{equation}
z_{\mathrm{ST}}=\frac{2L^{2}}{\lambda_{\mathrm{o}}}=\frac{2\alpha^{2}c^{2}}{\lambda_{\mathrm{o}}}T^{2}=\frac{T^{2}}{\pi|k_{2}|}.
\end{equation}
As the spatio-temporal lattice propagates freely, its structure first degrades under the influence of diffraction and dispersion [Fig.~\ref{fig:theory}(d) and Supplementary Movie~3 and Supplementary Movie~4], the lattice nodes spread out, and the periodic structure is obscured and becomes unrecognizable. Nevertheless, axial spatio-temporal revivals $\psi(x,\ell z_{\mathrm{ST}};t)=\psi(x,0;t-\ell\tfrac{z_{\mathrm{ST}}}{c})$ subsequently appear as shown in Fig.~\ref{fig:theory}(d), where $\ell$ is integer. The impact of the Janus form of $k_{z}$ for a V-wave is now clear in the exact matching of the Talbot lengths in space and time. It is crucial to recognize that the ST Talbot effect is \textit{not} the result of arranging for independent spatial and temporal Talbot effects to coincide. Instead, dispersion and diffraction proceed here in lockstep without external adjustments. Diffractive spatial spreading and dispersive temporal spreading are intrinsically synchronized, and a single Talbot length $z_{\mathrm{ST}}$ emerges. It is thus apt to call this phenomenon the ST Talbot effect.

Finally, the time-averaged intensity $I(x,z)=\int\!dt\,I(x,z;t)$, where $I(x,z;t)=|\psi(x,z;t)|^{2}$, as recorded by a detector lacking temporal resolution is (Supplementary):
\begin{equation}\label{Eq:TimeAveraged}
I(x,z)=\sum_{m}|\widetilde{\psi}_m|^2+\sum_{m}|\widetilde{\psi}_m\widetilde{\psi}_{-m}|\cos{(4\pi m\tfrac{x}{L}+\varphi_{m}-\varphi_{-m})},
\end{equation}
where $\widetilde{\psi}_{m}=|\widetilde{\psi}_{m}|e^{i\varphi_{m}}$ is the discretized spectral field amplitude at $k_{x}=m\tfrac{2\pi}{L}$. An unexpected effect emerges here: although the transverse spatial period of the spatio-temporal field lattice $I(x,z;t)$ is $L$ because $k_{x}=m\tfrac{2\pi}{L}$, Eq.~\ref{Eq:TimeAveraged} indicates that the transverse spatial period of the time-averaged intensity $I(x,z)$ is $L/2$ rather than $L$. Furthermore, despite the dramatic dynamical evolution of the lattice structure in $I(x,z;t)$ with free propagation [Fig.~\ref{fig:theory}(d)], this evolution is concealed in $I(x,z)$, which surprisingly lacks any axial dynamics as is clear from the absence of terms containing $z$ in Eq.~\ref{Eq:TimeAveraged}. This is another manifestation of the `veiled' Talbot effect (reported recently \cite{Yessenov20arXiv} in the spatial domain). A tight association of any kind between $k_{x}$ and $\omega$ masks temporal dynamics from having observable consequences in the time-averaged intensity, and also results in the halving of the time-averaged spatial period \cite{Yessenov20arXiv}. Because V-waves satisfy these necessary requirements (Eq.~\ref{Eq:VSpectrum}), the ST Talbot effect is also veiled after time-averaging.


\begin{figure*}[t!]
\centering 
\includegraphics[width=16cm]{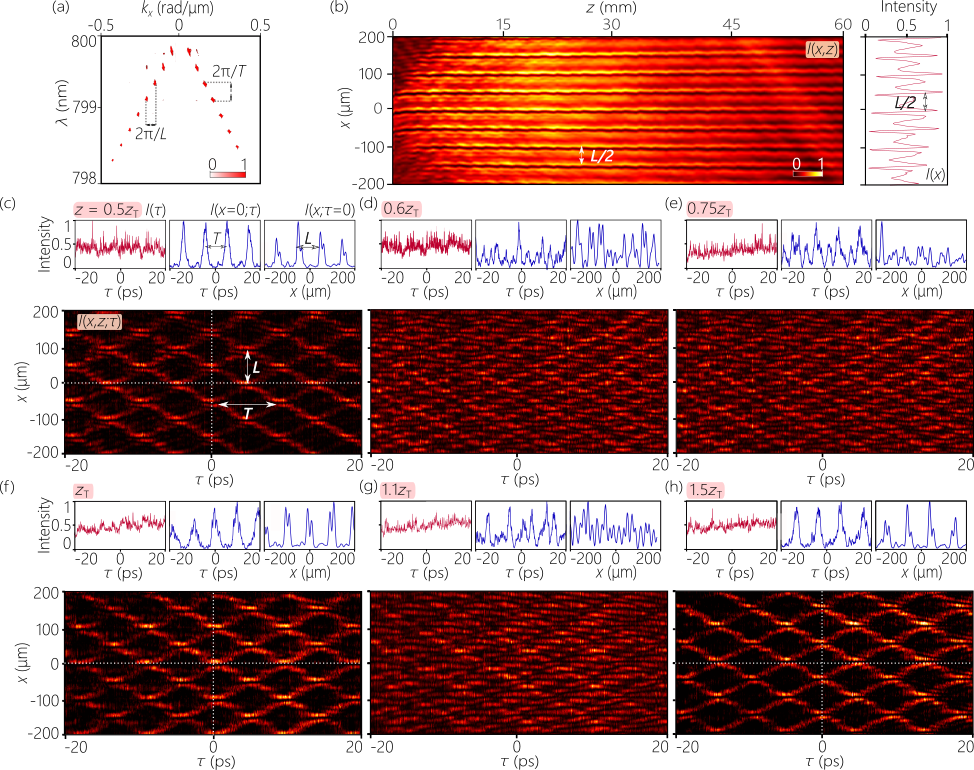} 
\caption{\textbf{Observation of the ST Talbot effect.} (a) Measured spatio-temporal spectrum with $\alpha=0.033$, $\lambda_{\mathrm{o}}=800$~nm, $\Delta\lambda=2$~nm, and discretized at multiples of $\tfrac{2\pi}{L}$ with $L=100$~$\mu$m, resulting in $T=10$~ps and $z_{\mathrm{ST}}=25$~mm. (b) The measured time-averaged intensity $I(x,z)$. The transverse period is $L/2=50$~$\mu$m rather than $L=100$~$\mu$m. Note the disparate length scales for $x$ and $z$. (c-h) Measured spatio-temporal intensity profiles $I(x,z;\tau)$ at $z=12.5$, 14, 19, 25, 27, and 37.5~mm, corresponding to $0.5z_{\mathrm{ST}}$, $0.6z_{\mathrm{ST}}$, $0.75z_{\mathrm{ST}}$, $z_{\mathrm{ST}}$, $1.1z_{\mathrm{ST}}$, and $1.5z_{\mathrm{ST}}$. The profiles are acquired in a reference frame moving at $c$. At each axial plane $z$, we plot the spatially averaged temporal profile $I(\tau)$, $I(x=0;\tau)$ and $I(x,\tau=0)$. The white dotted lines are guides for the eye and highlight the half-period shift in space and time at the full-Talbot plane ($z=z_{\mathrm{ST}}$) with respect to the mid-Talbot planes ($z=0.5z_{\mathrm{ST}}$ and $z=1.5z_{\mathrm{ST}}$). The corresponding theoretical predictions are provided in Supplementary Figure~3.}
\label{fig:experiment}
\end{figure*}

\section{Experimental results}

To experimentally validate this predicted ST Talbot effect, we prepare V-waves using the spatio-temporal wave-packet-shaping methodology developed in \cite{Yessenov19PRA,Kondakci19OL}. Starting with ultrafast plane-wave pulses, this pulse-shaper is designed to assign a spatial frequency $k_{x}$ to each wavelength $\lambda$ in the pulse spectrum according to the constraint in Eq.~\ref{Eq:VSpectrum}. In this configuration, utilizing a continuous spectrum yields a V-wave [Fig.~\ref{fig:theory}(a),(c)], while discretizing the spectrum helps realize the ST Talbot effect [Fig.~\ref{fig:theory}(b),(d)]. We make use of femtosecond pulses from a mode-locked Ti:sapphire laser, and sculpt the spatio-temporal spectrum via a spatial light modulator (SLM) over a temporal bandwidth $\Delta\lambda\approx2$~nm. The spatio-temporal spectrum projected onto the $(k_{x},\lambda)$-plane is acquired by a double Fourier transform in space and time, the time-averaged intensity $I(x,z)$ is obtained by scanning a CCD camera along the propagation axis $z$, and the spatio-temporal profiles $I(x,z;\tau)$ are measured at selected axial planes $z$ via interferometry with narrow reference pulses (Supplementary).

We first synthesize a continuous-spectrum V-wave with $\alpha=0.033$, corresponding to a GVD parameter of $|k_{2}|=1.3\times10^{6}$ fs$^2$/mm. The measurement results (Supplementary Figure~6) are in excellent agreement with the theoretical predictions plotted in Fig.~\ref{fig:theory}(c). We next introduce a periodic spatio-temporal lattice structure into the V-wave by discretizing the spatial spectrum at multiples of $\tfrac{2\pi}{L}$ with $L=100$~$\mu$m. By virtue of the linear relationship between $k_{x}$ and $\Omega$ (Eq.~\ref{Eq:VSpectrum}), discretizing $k_{x}$ naturally leads to a discretization of $\Omega$ at multiples of $\tfrac{2\pi}{T}$ (where $L$ and $T$ are related through Eq.~\ref{Eq:LandT}), as confirmed experimentally in Fig.~\ref{fig:experiment}(a). The temporal period is $T\approx10$~ps, corresponding to a pulse train of repetition rate $\approx100$~GHz and a ST Talbot length $z_{\mathrm{ST}}\approx25$~mm. The time-averaged intensity $I(x,z)$ in Fig.~\ref{fig:experiment}(b) shows a diffraction-free axial evolution independent of $z$ and with a transverse period $L/2=50$~$\mu$m, as expected from Eq.~\ref{Eq:TimeAveraged}.

Measurements of the spatio-temporal intensity profile $I(x,z;\tau)$ reveal an altogether different picture featuring complex propagation dynamics that is concealed in the time-averaged data. The profiles are acquired in a frame of reference propagating at $c$, and we plot $I(x,z;\tau)$ in Fig.~\ref{fig:experiment}(c)-(h) at selected axial planes extending over a full ST Talbot length $z_{\mathrm{ST}}$. In general, the revivals at the full- or mid-Talbot planes witness the reemergence of the temporal period $T$ and the transverse spatial period $L$. Clearly, the lattice profile at $0.5z_{\mathrm{ST}}$ is revived at $1.5z_{\mathrm{ST}}$. Furthermore, the lattice profile at $z_{\mathrm{ST}}$ is another revival, except that the spatial profile is shifted along $x$ by $L/2$ and along $\tau$ by $T/2$, such that the peaks in Fig.~\ref{fig:experiment}(f) correspond to minima in Fig.~\ref{fig:experiment}(c),(h) (and vice versa). In the intermediate planes [Fig.~\ref{fig:experiment}(d),(e),(g)], the distinct periodic features of the spatio-temporal lattice structure are lost, and the lattice disintegrates into an unrecognizable pattern after diffractive and dispersive spreading of the lattice nodes. Measured spatio-temporal intensity profiles acquired in other axial planes are assembled in Supplementary Movie~5, with corresponding calculated profiles $I(x,z;\tau)$ provided in Supplementary Figure~3. The excellent agreement between the measurements and calculations even in the fine-structure of the profiles is brought out by comparing the spatio-temporal profiles in closely separated planes (Supplementary Figure~7).

At the full- or mid-Talbot planes, any temporal slice at a fixed transverse position, $I(x=0,z;\tau)$ for example, reveals a periodic pulse train of period $T\approx10$~ps; and any spatial slice at a fixed delay, $I(x,z;\tau=0)$ for example, reveals a spatial periodic profile of period $L\approx100$~$\mu$m. The temporal structure is of course concealed in the time-averaged intensity $I(x,z)$ as shown in Fig.~\ref{fig:experiment}(b), but $I(x,z)$ remains periodic along $x$ (albeit with period $L/2$ rather than $L$). The spatial structure is in turn concealed in the space-averaged intensity $I(\tau,z)=\int\!dx\,I(x,z;\tau)$. However, no temporal structure is found in $I(\tau,z)$ at any $z$ [Fig.~\ref{fig:experiment}(c)-(h)]. This contrasting behavior between $I(x,z)$ and $I(\tau,z)$ is a consequence of the two-to-one relationship between $\pm k_{x}$ and $\omega$.

\begin{figure}[t!]
\centering 
\includegraphics[width=8.6cm]{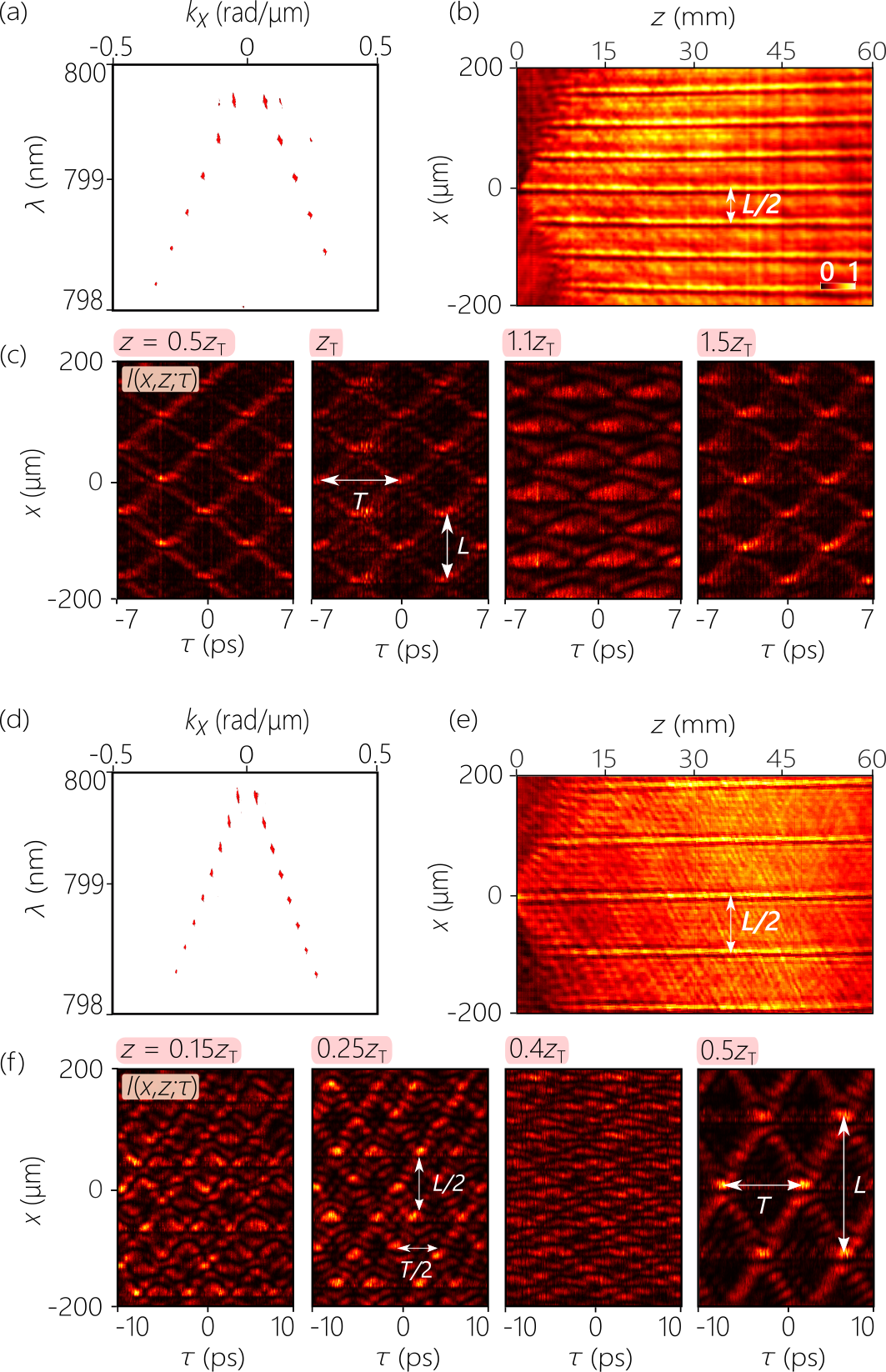} 
\caption{\textbf{Varying the spatial and temporal periods of the ST Talbot effect.} (a-c) The field parameters are $\alpha=0.048$, $\lambda_{\mathrm{o}}=800$~nm, $\Delta\lambda=2$~nm, and $L=100$~$\mu$m, resulting in $T=7$~ps and $z_{\mathrm{ST}}=25$~mm. (a) Measured spatio-temporal spectrum, (b) time-averaged intensity $I(x,z)$, and (c) spatio-temporal intensity profiles $I(x,z;\tau)$ at $z=12.5$, 25, 28, and 32.5~mm, corresponding to $z=0.5z_{\mathrm{ST}}$, $z_{\mathrm{ST}}$, $1.1z_{\mathrm{ST}}$ and $1.5z_{\mathrm{ST}}$. (d-f) The field parameters are $\alpha=0.067$, $\lambda_{\mathrm{o}}=800$~nm, $\Delta\lambda=2$~nm, and $L=200$~$\mu$m, corresponding to $T=10$~ps and $z_{\mathrm{ST}}=100$~mm. (d) Measured spatio-temporal spectrum, (e) time-averaged intensity $I(x,z)$, and (f) spatio-temporal intensity profiles $I(x,z;\tau)$ at $z=15$, 25, 40, and 50~mm, corresponding to $z=0.15z_{\mathrm{ST}}$, $0.25z_{\mathrm{ST}}$, $0.4z_{\mathrm{ST}}$ and $0.5z_{\mathrm{ST}}$. The intensity profiles in (c) and (f) are all acquired in a reference frame moving at $c$. The corresponding theoretical predictions are provided in Supplementary Figure~4.}
\label{fig:furtherExperiment}
\end{figure}

Further experimental explorations of the ST Talbot effect are shown in Fig.~\ref{fig:furtherExperiment}. We highlight here the surprising feature of V-waves whereupon varying a spatial or temporal parameter does not impact the exact balance between diffraction and dispersion lengths. First, we vary $\alpha$ while holding the spatial period $L$ fixed so that $z_{\mathrm{ST}}\approx25$~mm remains invariant. Changing the opening angle of the V-shaped spectrum results in a new discretization period along $\omega$ and hence to a new temporal period $T\approx7$~ps [Fig.~\ref{fig:furtherExperiment}(a)]. However, the change in $\alpha$ also modifies $|k_{2}|$ such that $\tfrac{T^{2}}{\pi|k_{2}|}$ remains equal to $\tfrac{L^{2}}{2\lambda_{\mathrm{o}}}$, and a single length scale still governs self-imaging in space and time. Although the time-averaged intensity $I(x,z)$ in Fig.~\ref{fig:furtherExperiment}(b) is \textit{identical} to that in Fig.~\ref{fig:experiment}(b), the underlying spatio-temporal evolution nevertheless reveals different lattice structure and dynamics [Fig.~\ref{fig:furtherExperiment}(c)]. 

We next increase the sampling rate along $k_{x}$ [Fig.~\ref{fig:furtherExperiment}(d)] to produce a larger spatial period $L=200$~$\mu$m and thus an increased ST Talbot length $z_{\mathrm{ST}}=100$~mm. By changing $\alpha$, we can retain the temporal period $T=10$~ps used in Fig.~\ref{fig:experiment}, while maintaining $\tfrac{T^{2}}{\pi|k_{2}|}=\tfrac{L^{2}}{2\lambda_{\mathrm{o}}}$. The measured transverse spatial period in the time-averaged intensity $I(x,z)$ is $L/2=100$~$\mu$m [Fig.~\ref{fig:furtherExperiment}(e)] rather than $L=200$~$\mu$m as revealed in the spatio-temporal intensity profiles $I(x,z;\tau)$ at the full- or mid-Talbot planes [Fig.~\ref{fig:furtherExperiment}(f)]. Whereas the field in the intermediate planes lacks any recognizable structure, the spatio-temporal lattice structure is revived at the quarter-Talbot planes $z=0.25z_{\mathrm{ST}}$, but at half the spatial \textit{and} temporal periods, $L/2$ and $T/2$, respectively. The measurements shown in Fig.~\ref{fig:furtherExperiment} are in excellent agreement with the corresponding calculated profiles in Supplementary Figure~4.

\section{Discussion}

The specific spatial-temporal coupling engendered by V-waves introduces novel qualitative and quantitative aspects into the propagation of optical fields. First, qualitatively, diffraction and dispersion are no longer physically independent phenomena. Rather, they are intertwined as a result of the Janus form of the propagation phase-factor $e^{ik_{z}z}$. Second, quantitatively, V-waves can produce extremely large values of GVD, rivaling or exceeding that of a Martinez compressor \cite{martinez87IEEE}. This dramatically reduces the dispersion length, leading to the first observation of an optical temporal Talbot effect in a freely propagating field unhampered by traditional restrictions imposed by material GVD. Uniquely, the diffraction and dispersion lengths are guaranteed to be identical for V-waves. The juxtaposition of these unique characteristics allows for the ST Talbot effect to be observed once the spectrum is discretized to impress a periodic spatio-temporal lattice structure upon the field. Indeed, the unambiguous observation of the ST Talbot effect is an independent confirmation of the exact matching of the diffraction and dispersion lengths, since even a few percent difference between them can diminish the predicted effect. Excitingly, taking our results here as a starting point, sculpting the spatio-temporal spectrum may lead to new optical field configurations in which the diffraction and dispersion lengths take on arbitrary values. Such a feature opens up prospects for studies of spatio-temporal solitons in nonlinear media by combining diffractive and dispersive effects at matching length scales.

Synthesizing optical fields such as V-waves endowed with precise spatio-temporal structure that restricts the spectrum to a reduced-dimensionality subspace has precedents. We have dubbed all pulsed fields in which each spatial frequency is associated with a single temporal frequency generically as ST wave packets \cite{Kondakci16OE,Parker16OE}. However, the emphasis to date has been on achieving propagation invariance \cite{Turunen10PO,FigueroaBook14}, which necessitates a linear relationship between $k_{z}$ and $\omega$ (rather than between $k_{x}$ and $\omega)$. The spectral support domain of such ST wave packets on the surface of the light-cone $k_{x}^{2}+k_{z}^{2}=(\tfrac{\omega}{c})^{2}$ lies at its intersection with a tilted plane that is parallel to the $k_{x}$-axis. The spectral projection onto the $(k_{z},\tfrac{\omega}{c})$-plane is a straight line, while that onto the $(k_{x},\tfrac{\omega}{c})$-plane is a conic section (Supplementary Figure~2). In such fields, the linear correlation between $k_{z}$ and $\omega$ introduces angular dispersion that combats diffractive spreading. Instances of such wave packets extend back to focus-wave modes \cite{Brittingham83JAP} and X-waves \cite{Lu92IEEEa,Saari97PRL}, among other examples \cite{Turunen10PO,FigueroaBook14}. Dispersion is absent from all such ST wave packets, which thus propagate rigidly in free space \cite{Kondakci17NP,Kondakci19NC}, transparent dielectrics \cite{Bhaduri19Optica,Bhaduri20NP}, planar waveguides \cite{Shiri20arxiv}, or as surface waves at planar interfaces \cite{Schepler20arxiv}. However, this absence of GVD renders such wave packets precludes them from realizing the ST Talbot effect. To the best of our knowledge, the V-waves examined here are the only example identified to date of an optical field with intrinsically equal diffraction and dispersion lengths, and is thus the unique platform for realizing the ST Talbot effect.

Finally, our theoretical results and measurements made use of optical waves. However, the same basic considerations apply to other physical waves, where it is a universal feature that the decoupled diffraction and dispersion length scales are incommensurate. The spatial Talbot effect has been realized in a variety of physical contexts besides optics, including acoustics \cite{Saiga85AO,Gao19Research}, atom matter waves \cite{Chapman95PRA}, rogue waves \cite{zhang15PRE}, water waves \cite{Bakman19AJP}, and elastic solid waves \cite{Berezovski14MRC} (and has been proposed for Bose-Einstein condensates \cite{Wen11APL,wen17SR}), whereas the temporal Talbot effect has been observed thus far in Bose-Einstein condensates \cite{Deng99PRL}. A combined spatio-temporal Talbot effect has \textit{not} been realized thus far in any physical wave. It is therefore an exciting prospect to explore approaches to the synthesis of V-waves in acoustics and matter-waves, which may then pave the way to observing the ST Talbot effect using such waves.

\section*{Funding}
Office of Naval Research (ONR) (N00014-17-1-2458 and N00014-20-1-2789).


\bibliography{diffraction}

\newpage

\end{document}